\documentclass[reviewcopy]{elsart}
\usepackage{color}
\usepackage{graphicx}
\usepackage[T1]{fontenc}
\usepackage{dcolumn}
\begin{document}
\begin{frontmatter}

\title{Synchronization and clustering in electroencephalographic  signals}
\author[first]{M. Escalona-Morán},
\ead{angele@ula.ve, Corresponding author}
\author[first]{M. G. Cosenza},
\author[third]{P. Guillén},
\author[fourth]{P. Coutin}
\address[first]{Centro de Física Fundamental,  Área de Caos y Sistemas Complejos,
 Universidad de Los Andes, Mérida, Venezuela}
\address[third]{Centro Nacional de Cálculo Científico, Universidad de Los Andes, Mérida,  Venezuela}
\address[fourth]{Departamento de Electrofisiología, Hospital San Juan de Dios, Mérida, Venezuela}

\begin{abstract}
The degree of synchronization and the amount of dynamical cluster
formation in electroencephalographic (EEG) signals are
characterized by employing two order parameters introduced in the
context of coupled chaotic systems subject to external noise.
These parameters are calculated in EEG signals from a group of
healthy subjects and a group of epileptic patients, including a
patient experiencing an epileptic crisis. The evolution of these
parameters shows the occurrence of intermittent synchronization and
clustering in the brain activity during an epileptic crisis.
Significantly,  the existence of an instantaneous maximum of
synchronization  previous to the onset of a crisis is revealed by
this procedure. The mean values of the order parameters and their
standard deviations are compared between both groups of
individuals.
\end{abstract}
\end{frontmatter}

\section{\label{sec1} Introduction}

In recent years there has been substantial interest in the
description of neural processes and electroencephalographic (EEG)
signals in the context of nonlinear dynamics and deterministic
chaos \cite{rapp,casdagli,babloyantz}.  Nonlinear dynamics opens
new windows for the understanding of EEG signals since the neural
activity involves nonlinear mechanisms at the microscopic level.
The complexity of the behavior of neural systems reflects the
effects of those underlying nonlinear mechanisms.  In particular,
the occurrence of synchronization in EEG signals associated to
various brain disorders, such as epilepsy and migraine, has been
the focus of much attention \cite{quyen1,angelini}.
Synchronization phenomena are expected to play a major role in
establishing the communication between different regions of the
brain \cite{gray,quyen2}.

The phenomenon of chaotic synchronization takes place when the
phase space trajectories of two or more coupled chaotic systems,
initially evolving on different attractors,  eventually converge
to a common trajectory \cite{fujisaka,pecora}. A related
phenomenon is dynamical clustering; it consists of the formation
of differentiated subsets of synchronized elements in a coupled
system \cite{kaneko}. Experimental observation of dynamical
clustering in coupled electrochemical systems has been reported
\cite{wang}.

This  article presents a  procedure that allows to characterize
the amount of synchronization and clustering occurring on  coupled
chaotic oscillators subject to common noise and applies these
concepts to EEG signals from healthy subjects and epileptic
patients. In Section \ref{sec2}, a theoretical model consisting of
a set of globally coupled chaotic R\"ossler oscillators and
subject to an external noise is introduced. Synchronization and
the formation of clusters, i.e., synchronized domains, under the
effect of noise are studied in this model. Two order parameters
are introduced in order to characterize  the synchronization and
the formation of clusters, i.e., synchronized domains, under the
effect of noise in this system. We calculate these order
parameters for different values of  the coupling strength between
the oscillators.  The data base of EEG signals used in this study
is described in Section \ref{sec3}; it consists of a group of ten
healthy subjects and a group of ten epileptic patients, including
one EEG signal of a patient experiencing an epileptic crisis. In
Section \ref{sec4}, we perform the analysis of synchronization and
cluster formation in the EEG signals from the data base. The
analysis yields quantitative indices that characterize the changes
of complexity in the brain activity of healthy and epileptic
subjects. Conclusions are presented in Section \ref{sec5}.

\section{\label{sec2} Globally coupled  R\"ossler  oscillators subject to external noise}
The Rössler equations constitute a paradigmatic dynamical model
that exhibits chaos. For some parameter values, these equations
give rise to a chaotic attractor possessing  fractal properties.
In order to introduce the concepts used in the characterization of
EGG signals, we consider a theoretical model consisting of system
of globally coupled Rössler oscillators and subject to an external
noise described by the following equations \cite{zanette}:
\begin{equation}
\label{eq1}
\begin{array}{rcl}
\dot{x}_{i} & = &-y_{i}-z_{i}+\varepsilon(\bar{x}-x_{i})+\xi_{i}(t) \\
\dot{y}_{i} & = &  x_{i}-ay_{i}+\varepsilon(\bar{y}-y_{i})  \\
\dot{z}_{i} & =& b-cz_{i}+x_{i}z_{i}+\varepsilon(\bar{z}-z_{i}) \, ,
\end{array}
\end{equation}
where $x_{i}$, $y_{i}$ and $z_{i}$ are the state variables of
oscillator  $i$;  with $ i=1, \ldots,N$; $N$ is the number of
oscillators; $\varepsilon$ represents the strength of the coupling
amount the oscillators; $\xi_{i}(t)$  is the source of noise
acting on oscillator $i$, with mean value
$\langle\xi_{i}(t)\rangle=0$ and correlation
$\langle\xi_{i}(t)\xi_{i}(t')\rangle=2S\delta(t-t')\delta_{ij}$; and $S$
is the amplitude of the noise. The mean value of the variables
$x_{i}$ is given by
\begin{equation}
\label{mean}
\bar{x}(t)=\frac{1}{N}\sum_{i=1}^{N} x_{i}(t),
\end{equation}
with similar expressions for $\bar{y}$  and $\bar{z}$.  The
addition of noise in Eq. (\ref{eq1}) is relevant because we wish
to establish a valid comparison between the behavior of this
dynamical model and that of physiological systems which usually
present noise and fluctuations,   as in the case of EEG signals.
The values of the parameters $a$, $b$ and $c$ employed here are
$a=b= 0.2$,  and $c=4.5$  for which the individual Rössler
oscillators are chaotic.

A cluster is defined as a subset of the population of oscillators
which are synchronized among themselves.  However, the presence of
noise in a system of interacting elements usually prevents perfect
synchronization.  In practice, we consider that a pair of elements
$i$ and  $j$ belong to a cluster if the distance $d_{ij}$ in phase
space between them, defined as
\begin{equation}
\label{dij}
d_{ij}=\sqrt{(x_{i}-x_{j})^{2}+(y_{i}-y_{j})^{2}+(z_{i}-z_{j})^{2}} \, ,
\end{equation}
is less than a threshold value $\gamma$, i.e., $d_{ij}<\gamma$.
The choice of $\gamma$ should be appropriate
for achieving differentiation between closely evolving clusters.

To characterize the amount of synchronization and the formation of
clusters in a system of coupled chaotic elements such as
Eq.(\ref{eq1}), two different order parameters can be introduced
\cite{zanette_2,manrubia}.  The first one,  $r(t)$, is defined as
the fraction of pairs of elements $(i,j)$  which are separated by
a distance $d_{ij}<\gamma$  at time $t$,
\begin{equation}
\label{r}
r(t)=\frac{1}{N(N-1)}\sum_{i=1}^{N}\sum_{j=1}^{N}\Theta[\gamma-d_{ij}(t)] \, ,
\end{equation}
where $\Theta(x)$ is the step function such that $\Theta(x)=0$ for
$x<0$ , and $\Theta(x)=1$ for $x\ge0$. The second order parameter,
$s(t)$, is the fraction of elements $i$  which at time $t$  have
at least  one other element $j$ located at a distance
$d_{ij}<\gamma$,
\begin{equation}\label{s}
s(t)=1-\frac{1}{N}\sum_{i=1}^{N}\prod_{j=1, j\ne i}^{N}\Theta[d_{ij}-\gamma] \,.
\end{equation}
The last term in Eq.(\ref{s}) is the fraction of elements whose
distance to any other element is greater than $\gamma$. Total
synchronization corresponds to $r(t)=s(t)=1$. On the other hand,
if all the population of oscillators is distributed in clusters,
we have $s(t)=1$ while $r(t)<1$, since the clusters may be
separated \cite{zanette}.  If some elements are in clusters but
others remain loose, then $r(t)<s(t)<1$.

System (\ref{eq1}) with $N=1000$ and noise of amplitude
$S=10^{-2}$ was integrated by using a fourth-order Runge-Kutta
method with time increments $\Delta t=10^{-2}$, starting from
random initial conditions. Noise was introduced through a reliable
random number generator.

\begin{figure}
\centerline{\hbox{
\includegraphics[scale=1,angle=0]{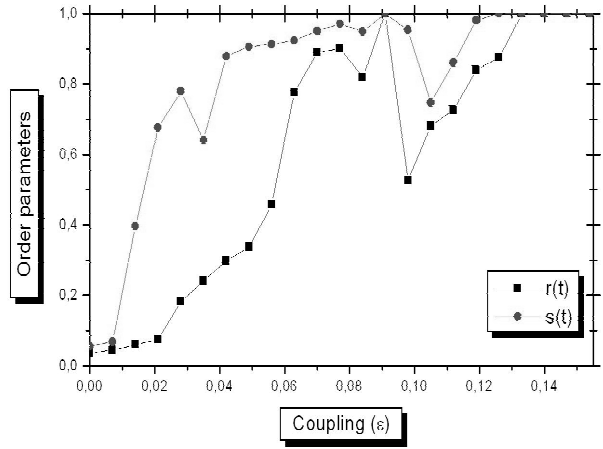}}}
\caption{ \label{fig:sys_synchro} Order parameters $r(t)$ and $s(t)$ as function of coupling $\varepsilon$, at time $t=5000$; with $N=1000$.}
\end{figure}

Figure~\ref{fig:sys_synchro} shows the order parameters  $r(t)$ and  $s(t)$ as functions of $\varepsilon$ at time $t=5000$. The threshold for cluster differentiation was set at $\gamma=10^{-3}$,
a value for which both $r(t)$ and  $s(t)$ reach statistically
stationary values.  For small values of $\varepsilon$, both $r(t)$
and $s(t)$ are less than $1$, indicating that not all oscillators
are forming clusters. In the interval $\varepsilon \in
(0.007,0.090)$,  $s(t)$ increases more rapidly than  $r(t)$; this
indicates that the oscillators tend to synchronize and to form
clusters. For $\varepsilon=0.091$, it is found $r(t)=s(t)=1$, and
therefore the elements are totally synchronized and forming a
single cluster. For $\varepsilon>0.133$ both order parameters
definitively reach their maximum values $r(t)=s(t)=1$ and the
system is totally synchronized.

\section{\label{sec3} Data base of EEG signals}
The EEG data base used in this study consist of records from $10$
healthy subjects and $10$ epileptic patients. One of the epileptic
records belongs to an individual who experienced a spontaneous
convulsive crisis during the EEG recording.  The record of the EEG
signal from each individual was done over $18$ channels connected
to scalp electrodes according to the international system $10-20$.
The potentials were measured with respect to a reference level
consisting of both ears short-circuited. The signal was
digitalized at a sampling frequency of $256$ Hz and $A/D$
conversion of $12$ bits,  and filtered to bandwidth between $0.5$
and $30$ Hz.

\begin{figure}
\centerline{\hbox{
\includegraphics[scale=1,angle=0]{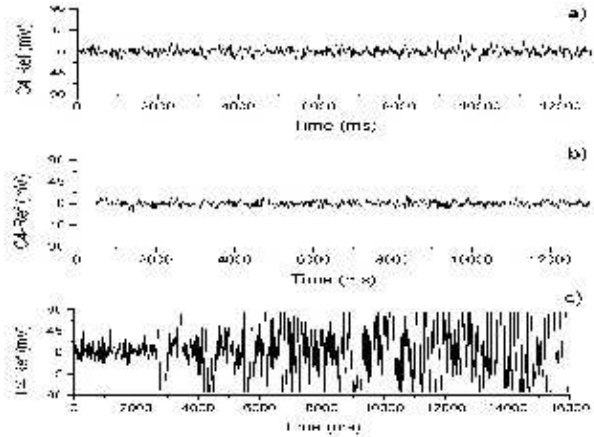}}}
\caption{\label{fig:eeg_all} (a) Channel C4 of the EEG signal from healthy subject. (b) Channel C4 of the EEG signal from an epileptic patient without crisis. (c) Channel C4 of the EEG signal from an epileptic patient experiencing a crisis. The flat extremes correspond to saturation of the electroencephalograph. }
\end{figure}

Figure \ref{fig:eeg_all} shows the records from the channel C4 of
the EEG signals corresponding to a healthy subject, an epileptic
patient without crisis, and an epileptic patient just before and
during a crisis, respectively.

\section{\label{sec4} Results}
In order to characterize the synchronization and cluster formation
in the EEG signals for each individual, we have employed the order
parameters $r(t)$ and $s(t)$ introduced in Sec. \ref{sec2}. At
time $t$ (ms), the values of $r(t)$ and $s(t)$ were calculated
for each EEG signal comprised of  $18$ channels.  The value of
$\gamma$ used corresponded to  $40\%$ of the standard deviation of
each EEG signal. This value of $\gamma$ is enough to discriminate
different clusters in the EEG signals.

\begin{figure}
\centerline{\hbox{
\includegraphics[scale=1,angle=0]{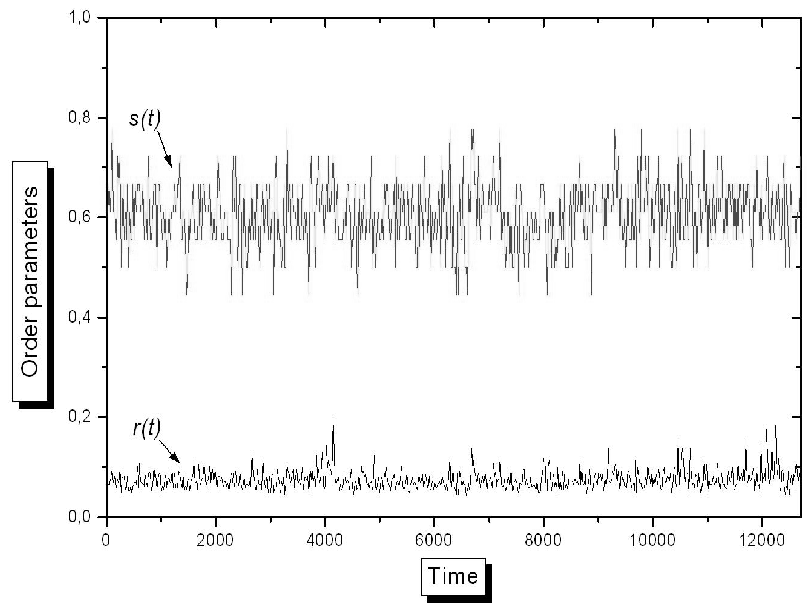}}}
\caption{ \label{fig:rs_healthy} Order parameters $r(t)$ and $s(t)$ for the EEG signal from a healthy subject.}
\end{figure}

Figure \ref{fig:rs_healthy} shows the order parameters $r(t)$ and
$s(t)$ for a healthy individual. The parameter $r(t)$ exhibits
relatively small values indicating desynchronization among the
different channels. The parameter$s(t)$ reaches higher values that
suggest the presence of some clusters. The maximum value reached
by $s(t)$ is $0.78$. The analysis of the EEG signals from the
others healthy subjects shows similar results to those shown in
Figure \ref{fig:rs_healthy}.

\begin{figure}
\centerline{\hbox{
\includegraphics[scale=1,angle=0]{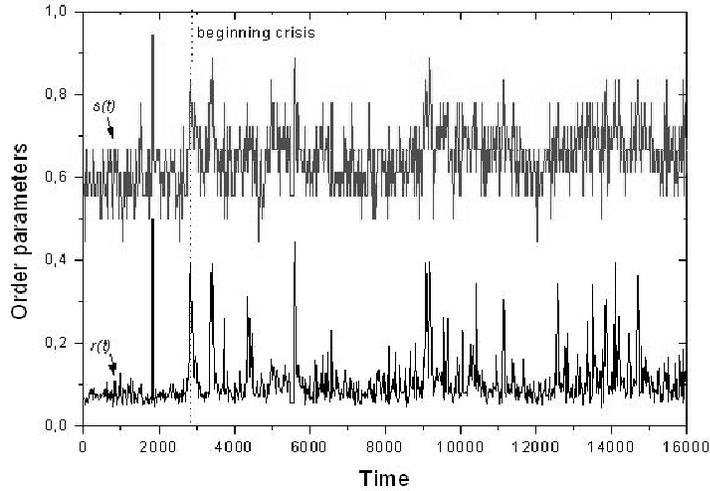}}}
\caption{ \label{fig:rs_crisis} Order parameters $r(t)$ and$s(t)$ for the EEG signal from the epileptic patient experiencing a crisis. The dotted vertical line indicates the onset of the crisis.}
\end{figure}

Figure \ref{fig:rs_crisis} shows the order parameters $r(t)$ and
$s(t)$ for a epileptic patient experiencing a spontaneous
epileptic crisis. The corresponding EEG signal is shown in Figure
\ref{fig:eeg_all}(c). The intervals where the amplifier was
saturated were excluded for the calculation of $r(t)$ and $s(t)$.
Note that the parameter $r(t)$ reaches a maximum value
$r(t)=0.51$ just two seconds before the beginning of the epileptic
crisis,  while at the onset of the crisis this parameter has a
value of  $r(t)=0.39$. Several maxima of $r(t)$ are observed
during the crisis. These maximum values  of $r(t)$ are greater
than those reached in EEG signals from healthy subjects.
Similarly, the parameter $s(t)$ in Figure \ref{fig:rs_crisis} also
exhibits a maximum value of $0.96$ at two seconds before the onset
of the crisis. In addition to this precursory maximum, the
parameter $s(t)$ shows several maxima during the epileptic crisis.
Most of these maxima correspond exactly to the maxima of the
parameter $r(t)$. This can be interpreted as the occurrence of
intermittent synchronization in the brain activity during an
epileptic crisis.

Figure \ref{fig:rs_crisis} reveals the occurrence of a synchronization maximum at two seconds prior
to the beginning of the epileptic crisis. However, this is not evident from the observation of the EEG
signal from the patient, shown in Figure \ref{fig:eeg_all}.c. This feature shows the potential that the
use of nonlinear dynamics techniques has in the analysis of physiological signals and how these techniques
can contribute to the prediction and diagnostics of some pathologies.

The parameters $r(t)$ and $s(t)$ calculated from the EEG signals from healthy subjects
and epileptic patients not experiencing crisis show a similar behavior. This might be due to the fact
that these patients were being treated with drugs.

\begin{table}[!h]
\caption{\label{tab:table} Mean values and standard deviations of parameters $r(t)$ and$s(t)$.}
\begin{tabular}{lllr}
\hline
\hline
 & Healthy Subjects & Epileptic patients & Epileptic patient\\
 & & without crisis & with crisis\\
\hline
$\overline{r}$ & 0.075 & 0.076 & 0.110\\
$\Delta r$ & 4.12$\times10^{-4}$& 2.62$\times10^{-4}$ & 4.06$\times10^{-3}$\\
$\overline{s}$& 0.607 & 0.604 & 0.646\\
$\Delta s$& 4.28$\times10^{-3}$ & 3.98$\times10^{-3}$ & 6.14$\times10^{-3}$\\
\hline
\hline
\end{tabular}
\end{table}

Table \ref{tab:table} displays the mean values $\overline{r}$,
$\overline{s}$, and standard deviations $\Delta r$ and $\Delta s$
of the parameters $r(t)$ and $s(t)$ over time and over the
population of healthy subjects, epileptic patients without crisis
at the time of recording and the epileptic patient experiencing a
crisis. In Table \ref{tab:table}, note that the mean value
$\overline{r}$ clearly differentiate an epileptic patient with
crisis from the other groups. The value of $\overline{r}$ does not
unambiguously distinguish between epileptic patients without
crisis and healthy subjects; however the value of the standard
deviation $\Delta r$ in the former group is about twice the value
corresponding to healthy subjects.

\section{\label{sec5} Conclusions}
Globally coupled systems, such as Eqs. ~(\ref{eq1}), constitute
theoretical models containing the essential ingredients that allow
the emergence of synchronization and the formation of clusters in
dynamical systems consisting of many interactive elements. The
central hypothesis of this article is that the brain can also be
described as a dynamical system formed by mutually coupled
nonlinear elements (neurons),  and therefore the collective
propierties of the brain are susceptible to be analyzed through
nonlinear dynamics techniques. With this aim, we have quantified
the amount of synchronization and clustering in EEG signals by
using order parameters previously introduced in the context of
coupled chaotic oscillators.

In the case of the epileptic patient experiencing a crisis, the order parameters $r(t)$ and $s(t)$  show
a series of instantaneous maxima in time which are absent in the evolution of those parameters in both healthy
subjects and epileptic patients without crisis. The maxima of $r(t)$ indicate instantaneous increases in synchronization.
During an epileptic crisis the mean value $\overline{r}$ is greater than those values corresponding to healthy subjects
and to epileptic patients without crisis. The standard deviation $\Delta r$ is an order of magnitude less in the
epileptic patient in crisis with respect to healthy subjects and to epileptic patients without crisis. These results
suggest that the epileptic crisis can be interpreted as a collective dynamical state characterized by the presence of
a high degree of synchronization between different zones of the brain.

A relevant finding of the present work is the presence of precursory maxima in both parameters $r(t)$ and $s(t)$
just before the onset of the epileptic crisis shown in Figure \ref{fig:rs_crisis}, while the corresponding
EEG signal does not show apparent changes at that moment.
Our results suggest that the use of nonlinear dynamics techniques as tools to characterize
EEG signals can contribute to a better understanding and diagnostic of some brain pathologies and,
as we have shown,  they may  help in the prediction of epileptic events. Nonlinear dynamics techniques
can also contribute to characterize other types of physiological signals.

\section*{Acknowledgments}
This work was supported by grant No. C-1285-04-05-A from Consejo de Desarrollo Científico,
Human\'{\i}stico y Tecnol\'ogico of Universidad de Los Andes, by CeCalCULA, and by Hospital San Juan de Dios,
M\'erida, Venezuela

\end{document}